\documentclass[twocolumn]{aastex631}

\usepackage{natbib}
\usepackage{newtxtext,newtxmath}
\usepackage{amsmath}
\usepackage{graphics,graphicx}
\usepackage{setspace}
\usepackage{caption,setspace}

\usepackage{caption}
\usepackage{subcaption}

\usepackage{mathtools}
\usepackage{enumitem}
\usepackage{lipsum}
\setlist[itemize]{leftmargin=*}

\usepackage{floatrow}
\usepackage[outercaption]{sidecap}    
\sidecaptionvpos{figure}{t}

\usepackage[T1]{fontenc}
\usepackage{ae,aecompl}
\pdfoutput=1

\shorttitle{Disappearing galaxies: the orientation dependence of JWST-bright, HST-dark, star-forming galaxy selection}
\shortauthors{R. K. Cochrane et al.}

\begin{document}
\title{Disappearing galaxies: the orientation dependence of JWST-bright, HST-dark, star-forming galaxy selection}

\author[0000-0001-8855-6107]{R. K. Cochrane}
\affiliation{Department of Astronomy, Columbia University, New York, NY 10027, USA}
\affiliation{Center for Computational Astrophysics, Flatiron Institute, 162 Fifth Avenue, New York, NY 10010, USA}
\footnote{email: rc3660@columbia.edu}

\author[0000-0001-5769-4945]{D. Angl\'es-Alc\'azar}
\affiliation{Department of Physics, University of Connecticut, 196 Auditorium Road, U-3046, Storrs, CT 06269-3046, USA\\}
\affiliation{Center for Computational Astrophysics, Flatiron Institute, 162 Fifth Avenue, New York, NY 10010, USA}

\author[0000-0002-3736-476X]{F. Cullen}
\affiliation{Institute for Astronomy, University of Edinburgh, Royal Observatory, Blackford Hill, Edinburgh, EH9 3HJ, UK}

\author[0000-0003-4073-3236]{C. C. Hayward}
\affiliation{Center for Computational Astrophysics, Flatiron Institute, 162 Fifth Avenue, New York, NY 10010, USA}

\begin{abstract}
\noindent Galaxies that are invisible in deep optical-NIR imaging but detected at longer wavelengths have been the focus of several recent observational studies, with speculation that they could constitute a substantial missing population and even dominate the cosmic star formation rate density at $z\gtrsim4$. The depths now achievable with JWST at the longest wavelengths probed by {\it{HST}}, coupled with the transformative resolution at longer wavelengths, are already enabling detailed, spatially-resolved characterisation of sources that were invisible to {\it{HST}}, often known as `{\it{HST}}-dark' galaxies. However, until now, there has been little theoretical work to compare against. We present the first simulation-based study of this population, using highly-resolved galaxies from the Feedback in Realistic Environments (FIRE) project, with multi-wavelength images along several lines of sight forward-modelled using radiative transfer. We naturally recover a population of modelled sources that meet commonly-used selection criteria ($H_{\rm{AB}}>27\,\rm{mag}$ and $H_{\rm{AB}}-\rm{F444W}>2.3$). These simulated {\it{HST}}-dark galaxies lie at high redshifts ($z=4-7$), have high levels of dust attenuation ($A_{V}=2-4$), and display compact recent star formation ($R_{1/2,\,\rm{4.4\,\mu\rm{m}}}\lesssim1\,\rm{kpc}$). Orientation is very important: for all but one of the 17 simulated galaxy snapshots with {\it{HST}}-dark sightlines, there exist other sightlines that do not meet the criteria. This result has important implications for comparisons between observations and models that do not resolve the detailed star-dust geometry, such as semi-analytic models or coarsely-resolved hydrodynamical simulations. Critically, we demonstrate that {\it{HST}}-dark sources are not an unexpected or exotic population, but a subset of high-redshift, highly-dust-attenuated sources viewed along certain lines of sight. 
\end{abstract}
\section{Introduction}
It has been known for decades that not all bright sub-millimeter galaxies (SMGs) have detectable optical/near-infrared (OIR) counterparts \citep{Lilly1999,Smail1999,Smail2000,Smail2002,Smail2004,Bertoldi2000,Frayer2000,Frayer2004,Dannerbauer2002,Wang2007a}. This can hamper redshift determination; indeed, the brightest source in the Hubble Deep Field at $850\,\mu\rm{m}$, HDF850.1, showed no clear OIR counterpart \citep{Hughes1998,Cowie2009}, and it took over a decade for its redshift to be identified \citep[$z=5.183$;][]{Walter2012}. \\
\indent Other populations of OIR-dark galaxies have been selected using optical-infrared colors. Galaxies first identified by red $R-K$ colors and faint $K-$band fluxes were called `Extremely Red Objects' \citep[EROs;][]{Elston1988,Hu1994}; some, but not all, of these sources were also identified in the sub-millimeter \citep{Cimatti1998,Dey1999}. Later, {\it Spitzer}-IRAC imaging \citep{Fazio2004a} enabled efficient surveys of new red populations \citep{Wilson2004,Huang2011} and better separation of passive and star-forming galaxies. \cite{Wang2016} introduced a color selection to identify massive, high-redshift galaxies using a combination of deep {\it{HST}} $H-$band imaging and IRAC CH2: $H_{\rm{AB}}-[4.5]>2.25\,\rm{mag}$. $62\%$ of sources selected in this way were detected down to $S_{870\,\mu\rm{m}}>0.6\,\rm{mJy}$ \citep{Wang2019a}. OIR-faint sources have also been detected at radio wavelengths \citep{Kondapally2021,Talia2021,Enia2022,VanderVlugt2022,Behiri2023}. \\
\indent OIR-faint galaxies are important for several reasons. Firstly, several works have suggested that they may contribute significantly to the cosmic star formation rate density (SFRD) at $z\gtrsim3$, though estimates of the contribution vary (e.g. \citealt{Wang2016} calculated that $H-$dark sources contribute $15-25\,\%$ of the SFRD at $z=4-5$, while \citealt{Sun2021a} estimated $8^{+8}_{-4}\,\%$ at $z=3-5$; see also \citealt{Williams2019a,Yamaguchi2019,Xiao2023}). Hence, excluding OIR-faint sources from censuses of star formation (e.g. by using $H-$ or $K-$selected samples, or by requiring OIR counterparts for sources selected in other ways) will lead to a redshift-dependent underestimate of the SFRD. Secondly, degeneracies between the signatures of redshift, age and dust attenuation in photometry result in the possibility of dusty sources contaminating surveys of high-redshift quiescent galaxies \citep[e.g.][]{Smail2002a,Dunlop2007,Simpson2017}. Thirdly, OIR-faint sources may help constrain theoretical models of galaxy formation and evolution. \cite{Wang2019a} showed that the number density of $H_{\rm{AB}}>27\,\rm{mag}$ sources predicted by an early version of the {\sc{l-galaxies}} semi-analytic model (SAM) \citep{Henriques2015} lies $\sim2$ orders of magnitude below the observed number density. They speculated that our understanding of massive-galaxy formation may require substantial revision. \\
\indent The advent of ALMA has facilitated statistical studies of large samples of sub-millimeter bright galaxies, including locating and resolving sources identified with single dish telescopes \citep[e.g.][]{Hodge2013,Karim2013,Simpson2014,Simpson2015a,Simpson2020a,Stach2018}. This has enabled more detailed exploration of the number density and physical properties of the subset of sub-millimeter-bright galaxies that are also OIR-dark. Using ALMA follow-up of SCUBA-2-identified sources within the UKIDSS Ultra Deep Survey \citep{Simpson2016,Stach2018,Stach2019,Dudzeviciute2019}, \cite{Smail2020} found that $15\pm2$\,\% of the bright ($S_{850\,\mu\rm{m}}>3.6\,\rm{mJy}$) SMG population is fainter than $K_{\rm{AB}}=25.3\,\rm{mag}$. These $K-$faint galaxies tend to have higher redshifts and dust attenuation than their $K-$detected counterparts (see also the high photometric redshifts derived for the $HST-$dark, $1.1\,\rm{mm}$-selected sources from \citealt{Franco2018}, and for OIR-dark galaxies selected at longer wavelengths: \citealt{Williams2019a,Casey2021a,Manning2022}). However, given the relatively low angular resolution of IRAC ($\rm{FWHM}\sim2''$) and the typically small angular sizes of OIR-dark sources, characterising the spatial distribution of rest-frame optical emission was not possible until the launch of JWST. \\
\indent JWST has provided the sensitivity and wavelength coverage required to gather high angular resolution (compared to IRAC), multi-wavelength imaging and hence constrain the properties of sources that were invisible to {\it{HST}} \citep[e.g.][]{Fujimoto2023,Gomez-Guijarro2023,Perez-Gonzalez2022,Rodighiero2023,Smail2023}. Employing a similar selection used in previous studies to identify dusty galaxies at $z=3-6$ ($H_{\rm{AB}}>27$ \& $H_{\rm{AB}}-\rm{F444W_{AB}}>2.3$; \citealt{Caputi2012,Wang2016}), \cite{Barrufet2022} selected a sample of 33 {\it{HST}}-dark sources within CEERS NIRCam imaging \citep{Finkelstein2022}. They showed that galaxies selected using these criteria tend to be fairly dust-obscured ($A_{V}\sim2$), massive ($M_{\star}\sim10^{9}-10^{10.5}\,\rm{M_{\odot}}$) star-forming galaxies at $z\sim2-8$ that lie approximately along the main sequence. Using the same data, \cite{Nelson2022} identified a population of spatially extended ($R_{e,\,4.4\,\mu\rm{m}}>0.17''$), $4.4\,\mu\rm{m}$-bright, $H-$faint sources, with higher inferred stellar masses, the majority in the range $M_{\star}=10^{10-11}\,\rm{M_{\odot}}$ (using EAzY, the median stellar mass is $M_{\star}=10^{10.5}\,\rm{M_{\odot}}$, but they also find systematic differences between SED fitting codes). These $4.4\,\mu\rm{m}$-identified sources appear to have lower stellar masses, on average, than sub-millimeter-identified OIR-dark sources. For example, the median stellar masses of the $K$-faint and whole sample of \citealt{Smail2020} are $M_{\star}=10^{11.10\pm0.04}\,\rm{M_{\odot}}$ and $M_{\star}=10^{11.00\pm0.06}\,\rm{M_{\odot}}$, respectively, and some of the most extreme sub-millimeter bright sources are likely even more massive (e.g. the stellar mass derived for the $S_{850\,\mu\rm{m}}=15.3\pm0.4\,\rm{mJy}$, $z=4.26$ source presented by \citealt{Smail2023} is $10^{11.8}\,\rm{M_{\odot}}$). \\
\indent Despite substantial observational interest in {\it{HST}}-dark galaxies, until now there have been no focused theoretical studies on the nature of {\it{HST}}-dark galaxies. In this paper, we study the physical properties of galaxies modelled with highly-resolved zoom-in simulations using the Feedback in Realistic Environments (FIRE; \citealt{Hopkins2014,Hopkins2017,Hopkins2022}) model that meet observational {\it{HST}}-dark selection criteria. In Section \ref{sec:sims}, we describe the simulations and the radiative transfer calculations performed to forward-model observable emission. We describe the selection of {\it{HST}}-dark galaxies. In Section \ref{sec:HST-dark_nature}, we study the demographics, dust attenuation properties and sizes of our simulated {\it{HST}}-dark galaxies, and highlight a substantial line-of-sight dependence of the source section. We draw our conclusions in Section \ref{sec:conclusions}.

\section{Simulations and HST-dark source selection}\label{sec:sims}
\subsection{FIRE simulations}\label{sec:FIRE_sims}
The FIRE\footnote{\url{https://fire.northwestern.edu}} project is a suite of state-of-the-art hydrodynamical cosmological zoom-in simulations described fully in \citet[][FIRE-1]{Hopkins2014}, \citet[][FIRE-2]{Hopkins2017} and \citet[][FIRE-3]{Hopkins2022}. In this paper, we study galaxies modelled with FIRE-2, which uses the ``meshless finite mass'' mode of the $N$-body+hydrodynamics code GIZMO\footnote{\url{http://www.tapir.caltech.edu/~phopkins/Site/GIZMO.html}} \citep{hopkins2015}; gravitational forces are computed following the methods presented in \cite{Hopkins2013}, using an improved version of the parallel TreeSPH code GADGET-3 \citep{Springel2005}. Cooling and heating processes including  free-free, photoionization/recombination, Compton, photo-electric, metal-line, molecular and fine-structure processes are modelled from $T=10\,\rm{K}$ to $T=10^{10}\,\rm{K}$. Star particles form from locally self-gravitating, molecular, Jeans unstable gas above a minimum hydrogen number density $n_{H}\geq1000\,\rm{cm}^{-3}$. Each star particle represents a single stellar population with known mass, age, and metallicity, injecting feedback locally in the form of mass, momentum, energy, and metals from Type Ia and Type II Supernovae (SNe), stellar winds, photoionization and photoelectric heating, and radiation pressure, with all feedback quantities and their time dependence taken directly from the {\sc{starburst}99} population synthesis model \citep{Leitherer1999}. \\
\indent We study the central galaxies of eight massive haloes originally selected and simulated by \citet{Feldmann2016, Feldmann2017a} as part of the {\sc MassiveFIRE-1} suite. The same haloes were studied in \cite{Cochrane2022} and \cite{Cochrane2023a}. The first four haloes are drawn from the `A-series' (A1, A2, A4 and A8); these haloes were selected to have dark matter halo masses of $M_{\rm{halo}}\sim10^{12.5}\,\rm{M_{\odot}}$ at $z=2$. The `A-series' halos studied in this paper are drawn from \cite{Angles-Alcazar2017}, who re-simulated them down to $z=1$ with the upgraded FIRE-2 physics model \citep{Hopkins2017}. We supplement these haloes by re-running four more haloes from \cite{Feldmann2017a}, with the updated FIRE-2 physics. Two of the haloes are drawn from their `B-series' (B1 and B2) and two from the `C-series' (Cm1:0, hereafter C1, and C2:0, hereafter C2). Haloes B2, C1 and C2 tend to be more massive than the A-series haloes. The mass resolution for both gas and star particles is $3.3\times10^4\,\rm{M_{\odot}}$, $2.7\times10^5\,\rm{M_{\odot}}$, and $2.2\times10^6\,\rm{M_{\odot}}$, for A-, B-, and C-series haloes, respectively. For dark matter particles, the respective mass resolutions are: $1.7\times10^5\,\rm{M_{\odot}}$, $1.4\times10^6\,\rm{M_{\odot}}$, and $1.1\times10^7\,\rm{M_{\odot}}$. Convergence tests for these simulations are presented in \cite{Cochrane2022} (see Appendix B). 

\subsection{Modelling observable emission}\label{sec:FIRE_RT}
We model spectral energy distributions (SEDs) and multi-wavelength emission maps at every snapshot, between $z\sim8$ and $z\sim1$, for each of the eight simulated halos (in total, $1736$ snapshots). Following \cite{Cochrane2019,Cochrane2022,Cochrane2023a,Cochrane2023b} we use the {\sc{skirt}}\footnote{\url{http://www.skirt.ugent.be}} radiative transfer code \citep[version 8;][]{Baes2011,Camps2014} to make predictions for emission between rest-frame ultraviolet (UV) and far-infrared (FIR) wavelengths along seven lines of sight. For all snapshots, gas and star particles within $0.1R_{\rm{vir}}$ are drawn directly from FIRE-2 simulation data. Dust particles are assumed to follow the distribution of the gas particles, with a dust-to-metals mass ratio of $0.4$ \citep{Dwek1998,James2002}. This is a reasonable assumption for enriched, massive galaxies like these. We have checked that using the metallicity-dependent dust-to-metal ratios draw from the relations derived by \cite{Remy-Ruyer2014a} would yield similar values. We assume dust destruction at $>10^6\,{\rm K}$ \citep{Draine1979,Tielens1994}. Following \cite{Cochrane2019,Cochrane2022,Cochrane2023a,Cochrane2023b}, we model a mixture of graphite, silicate and PAH grains using the \cite{Weingartner2001} Milky Way dust prescription. Star particles are assigned \citet{Charlot2003} SEDs based on their ages and metallicities. We perform the radiative transfer on an octree dust grid, in which cell sizes are adjusted according to the dust density distribution, with the condition that no dust cell may contain more than $0.0001\%$ of the total dust mass of the galaxy. {\sc{skirt}} parameter convergence tests are presented in \cite{Cochrane2022}. \\
\indent The output from {\sc{skirt}} comprises predictions for global galaxy SEDs as well as maps of the resolved emission at each of the $\sim100$ wavelengths modelled, for seven lines of sight. We transform flux densities in Jy to AB magnitudes (using a zero-point of $3631\,\rm{Jy}$).

\subsection{HST-dark source selection}\label{sec:HST-dark_selection}

Following \cite{Barrufet2022}, we select sightlines from our snapshots that meet the following $H-$band and color criteria:
\begin{equation}
    H_{\rm{AB}}>27\,\rm{mag}
\end{equation}
\hspace{3.9cm} {\it{and}}
\begin{equation}
    H_{\rm{AB}}-\rm{F444W}_{\rm{AB}}>2.3.
\end{equation}

\begin{figure}
\includegraphics[width=\columnwidth]{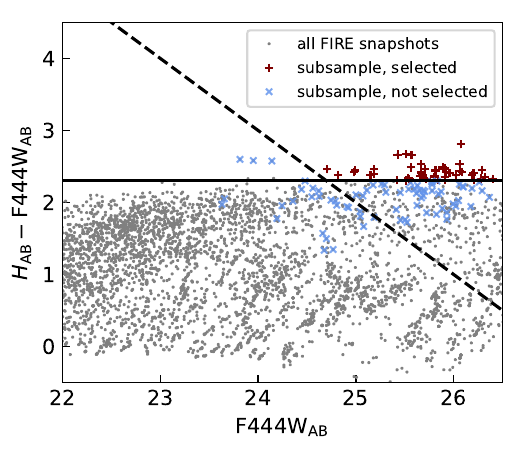}
\caption{Selection of snapshot orientations detected as {\it{HST}}-dark (dark red), according to the criteria: $H_{\rm{AB}}-\rm{F444W}_{\rm{AB}}>2.3$ (solid line) {\it{and}} $H_{\rm{AB}}>27\,\rm{mag}$ (dashed line). Lines of sight that do not meet this criteria, when one or more lines of sight of the same galaxy snapshot do, are plotted in blue. These tend to be slightly too blue, slightly too $H-$bright, or both. Lines of sight for snapshots with no orientations meeting the criteria are plotted in gray.\label{fig:color-mag}}
\end{figure}

Variations in both $H$-band magnitude and color with orientation result in snapshots for which some lines of sight meet the criteria and some do not. We define two subsets of lines of sight of the {\it{HST}}-dark subsample: `selected' are orientations that meet the criteria; `not-selected' are orientations that do not meet the criteria, where there exist other orientations of the same snapshot (i.e. the same galaxy and redshift) that do meet the criteria. In total, we obtain $17$ independent galaxy snapshots that meet the selection criteria at one or more viewing angles. We simulate a total of $119$ viewing angles of these sometimes-{\it{HST}}-dark snapshots, where $46$ meet the selection and $73$ do not. We study the viewing angle dependence in more detail in Section \ref{sec:model_comparison}. \\
\indent In Figure \ref{fig:color-mag}, we show the positions in the color-magnitude plane of selected and not-selected lines of sight of snapshots that are classified as {\it{HST}}-dark from one or more viewing angles. In gray, we show all seven lines of sight of FIRE snapshots with no {\it{HST}}-dark sightlines. The selected {\it{HST}}-dark subsample occupies a similar region of this parameter space to the observed sample of \cite{Barrufet2022}, giving confidence in the realism of our simulated galaxies. Lines of sight in the not-selected sample tend to lie just outside the boundaries of the selection, either because they are slightly too bright at $1.6\,\mu\rm{m}$, slightly too blue, or both. Only one simulated galaxy snapshot is {\it{HST}}-dark from all seven modelled viewing angles. This implies that {\it{HST}}-dark sources selected in this way are not an entirely unique population; instead, they are sources viewed from preferential orientations that would be visible by {\it{HST}} if viewed from other directions. In Section \ref{sec:HST-dark_nature}, we explore the physical properties of the selected {\it{HST}}-dark sources and the drivers of this orientation dependence.

\subsection{Comments on selection criteria}\label{sec:HST-dark_selection_criteria}
Here, we comment on the effects of each of the selection criteria on the demographics of the selected population. The $H-$dark criterion favours the selection of higher redshift sources, whereas the color criterion favours intermediate redshifts. A looser color criterion (i.e. extending to bluer colors) would lead to the inclusion of higher redshift $H-$faint sources into the sample. All of our selected snapshots lie within the redshift range $4<z<7$, but the redshift distribution and range will depend on the details of the mass assembly and metal enrichment of the simulated galaxies. While a modest-sized sample constructed in this way does not allow predictions for the redshift distribution of observed sources, we can note some general trends. Firstly, the selection criteria identify simulated sources within the general redshift range of the {\it{HST}}-dark, IRAC/F444W-bright sources reported by observational studies \citep[e.g.][]{Wang2019a,Barrufet2022,Nelson2022}. The modelled $4.4\,\mu\rm{m}$ emission of our sources is fairly faint ($\rm{F444W_{AB}} \gtrsim 24.8\,\rm{mag}$) compared to observed {\it{HST}}-dark sources, but still within the range measured by \cite{Barrufet2022}.\\
\indent Importantly, the OIR-dark, sub-millimeter-bright population tends to extend to lower redshifts: the median photometric redshift of the $K-$dark SMGs studied by \cite{Smail2020} is $z=3.44\pm0.06$, the majority lying in the range $z=3-4$, with a handful below $z=3$. Our simulated sources become too $H-$bright below $z\sim4$, but we would expect that dustier, more sub-millimeter-bright sources would be more highly attenuated in the $H-$band and hence meet the $H-$faint criterion (the degeneracy between $A_{V}$ and $z$ is also discussed in \citealt{Caputi2012,Wang2016,Sun2021a}). Due to the limited dynamic range of halos simulated here, we do not model these sub-millimeter-bright, highly attenuated sources. Instead, we focus on the sub-millimeter-faint ($S_{850\,\mu\rm{m}}\lesssim1\,\rm{mJy}$) HST-dark population. \\
\indent Finally, we comment on some modelling choices that affect the predicted fluxes of our simulated galaxies, and therefore the selection. In this work, following the methods of \cite{Cochrane2019,Cochrane2022,Cochrane2023a,Cochrane2023b}, we do not implement any models for unresolved, small-scale dust. \cite{Ma2019} argues that the FIRE galaxies are sufficiently well-resolved that this is not necessary. Tests have, however, been performed using the MAPPINGS III models \citep{Groves2008} to model emission from the warm dust associated with the unresolved birth clouds of young star clusters. \cite{Liang2020} studied the impact of varying two MAPPINGS parameters, $\log C$ (the HII region compactness) and $f_{\rm{PDR}}$ (the covering fraction of the associated photo-dissociation regions) on the predicted SED. They found that the choice of $\log C$ had little impact on the UV-optical SED, while a higher $f_{\rm{PDR}}$ yields a higher dust optical depth, and therefore reduced UV-optical fluxes and a steeper dust attenuation curve. Although the effects on the SED are modest, we note that implementing the MAPPINGS models with a non-zero $f_{\rm{PDR}}$ could result in more snapshots meeting the {\it{HST}}-dark criteria (due to both lower $H-$band fluxes and redder colours).

\begin{figure*} 
\includegraphics[width=0.33\columnwidth]{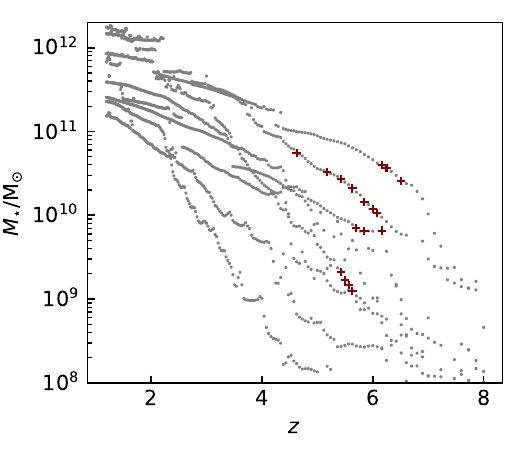}
\includegraphics[width=0.33\columnwidth]{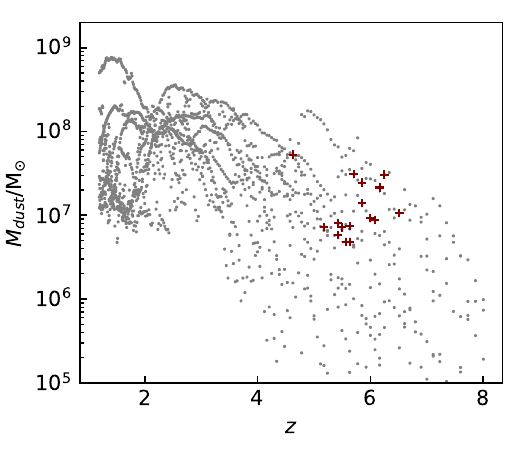}
\includegraphics[width=0.33\columnwidth]{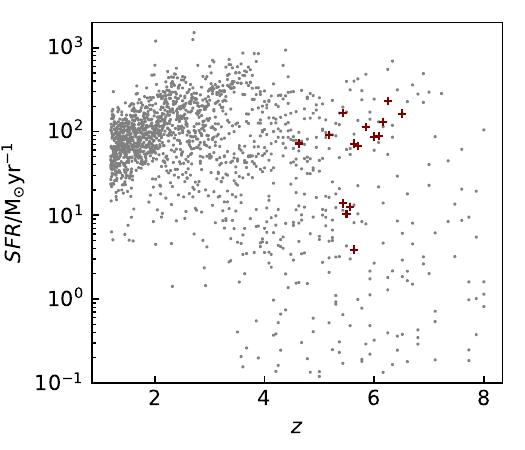}
\caption{The redshift evolution of stellar mass (top left), dust mass (center) and star formation rate (right), for all eight FIRE halos in this study (grey). Snapshots with {\it{HST}}-dark sightlines are plotted in red. Simulated galaxies with {\it{HST}}-dark sightlines span redshifts $\sim4-7$, with a broad range of stellar masses (from $\sim10^{9}\,\rm{M_{\odot}}$ to $\sim10^{11}\,\rm{M_{\odot}}$).\vspace{0.1cm}}
\label{fig:demographics}
\end{figure*}

\section{The nature of HST-dark sources}\label{sec:HST-dark_nature}

\subsection{HST-dark source demographics}\label{sec:HST-dark_demographics}
In Figure \ref{fig:demographics}, we show the redshift evolution of stellar mass, dust mass and star formation rate for the eight FIRE halos studied. We briefly summarise the demographics of snapshots with {\it{HST}}-dark lines of sight here. Stellar masses of the galaxies in selected snapshots range from $\sim10^{9}\,\rm{M_{\odot}}$ to $\sim10^{11}\,\rm{M_{\odot}}$, with mean $\log_{10}(M_{\star}/\rm{M_{\odot}})=9.8$. Dust masses range from $\sim5\times10^{6}\,\rm{M_{\odot}}$ to $\sim5\times10^{7}\,\rm{M_{\odot}}$, with mean $\log_{10}(M_{\rm{dust}}/\rm{M_{\odot}})=7.0$. Star formation rates (defined here as instantaneous) range from a few to a few hundred solar masses per year, with mean $\rm{SFR}=56\,\rm{M_{\odot}\rm{yr}^{-1}}$. These are in broad agreement with the inferred physical properties of $4.4/4.5\,\mu\rm{m}$-selected {\it{HST}}-dark sources \citep{Sun2021a,Barrufet2022,Nelson2022}. In contrast, the sub-millimeter-selected $K-$dark galaxies studied by \cite{Smail2020} have higher derived dust masses ($>10^{8}\,\rm{M_{\odot}}$), with stellar masses $\sim10^{11}\,\rm{M_{\odot}}$; the majority of those sources lie at $z<4$.

\begin{figure*} 
\includegraphics[scale=0.82]{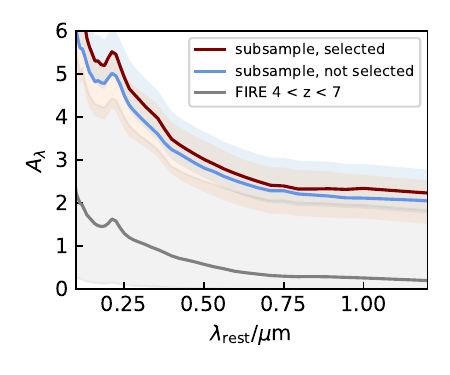}
\hspace{-0.49cm}
\includegraphics[scale=0.82]{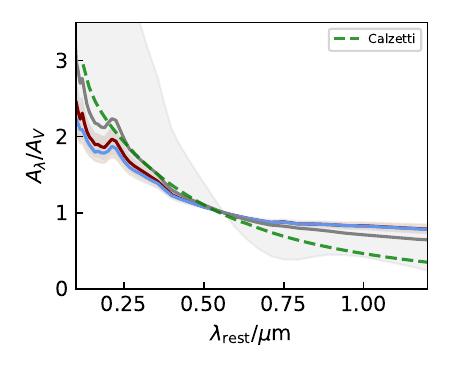}
\hspace{-0.49cm}
\includegraphics[scale=0.81]{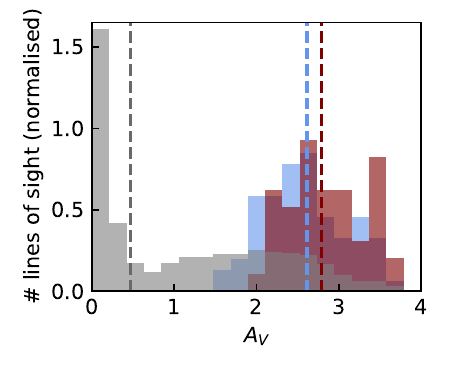}
\caption{Left: dust attenuation curves for our subsamples of selected and not-selected sightlines (red and blue) of {\it HST}-dark sources, compared to those of all FIRE-2 snapshots at $4<z<7$ (gray). Solid lines and shaded regions show median curves and $1\,\sigma$ range for all lines of sight that fall into a given subsample, respectively. Center: dust attenuation curves as in the left-hand panel, now normalised by $A_{V}$. We overplot the canonical \cite{Calzetti2000} law. In both panels, the average attenuation curves of selected and not-selected lines of sight are similar, with the median not-selected curve lying  below the selected curve by $\Delta A_{V} < 0.22$ at all wavelengths. Right: the distribution of $A_{V}$ values for each of the samples, with dashed lines showing medians (median $A_{V}=2.8,\,2.6$ for selected and not-selected orientations, respectively, compared to $A_{V}=0.5$ for the whole FIRE sample at $z=4-7$). Snapshots that are selected as {\it{HST}}-dark along one or more lines of sight are distinguished from the overall population by particularly high $A_{V}$ values.\\}
\label{fig:attenuation_curves}
\end{figure*}

\subsection{Dust attenuation of HST-dark sources}\label{sec:dust_attenuation_law}
By comparing intrinsic and attenuated emission as a function of wavelength, we can study the effective attenuation curves of the simulated galaxies at each snapshot. In Figure \ref{fig:attenuation_curves}, we plot median dust attenuation $A_{\lambda}$ (left) and median $V$-band-normalised dust attenuation $A_{\lambda}/A_{V}$ (center) versus wavelength, for the subsample of lines of sight selected as {\it{HST}}-dark and red (red line) and non-selected lines of sight of the same snapshots (blue line). For comparison, we also show curves generated using all snapshots in the redshift range within which our {\it{HST}}-dark-selected sources fall, $4<z<7$. It is clear from the left-hand panel that both selected and not-selected lines of sight of the {\it{HST}}-dark snapshots are significantly more attenuated, on average, than other sources in the same redshift range. Once normalised by $A_{V}$, though, the shapes of the attenuation curves are very similar (center panel) and grayer than the curve derived by \cite{Calzetti2000} (note, though, that we have used the same dust grain model for all simulated sources and snapshots: different choices of dust grain mix and size distributions would change the details of the attenuation and possibly also change the numbers of simulated sources that meet the {\it{HST-}}dark selection criteria). In the right-hand panel, we show a histogram of $A_{V}$ for the different populations. The median $A_{V}$ values for selected and not-selected lines of sight of {\it{HST-}}dark snapshots are $A_{V}=2.8$ and $2.6$, respectively, while $A_{V}=0.5$ for FIRE snapshots in the redshift range $z=4-7$.

\begin{figure*}
\begin{subfigure}[b]{1.0\textwidth}
\includegraphics[scale=0.67]{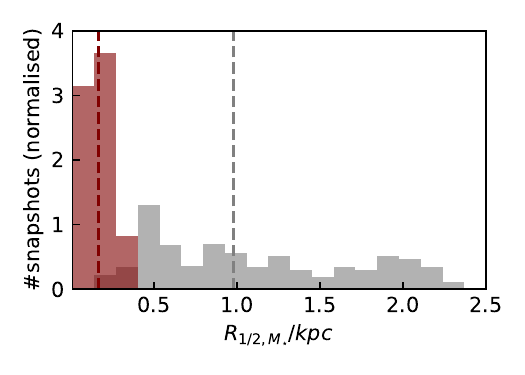}
\includegraphics[scale=0.67]{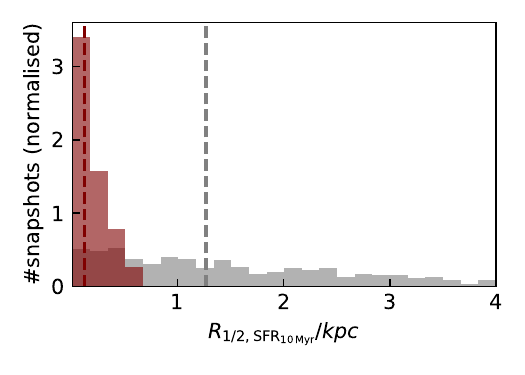}
\includegraphics[scale=0.67]{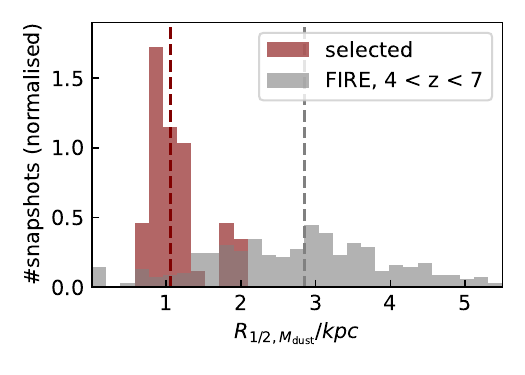}
\caption{Distributions of radii containing half the stellar mass (left), SFR over the last $10\,\rm{Myr}$ (center) and dust mass (right), for sources with {\it HST-}dark sightlines (red) compared to all FIRE-2 snapshots in the redshift range $4<z<7$ (grey). Snapshots with {\it HST-}dark  sightlines display extremely compact stellar mass, star formation ($R_{1/2,\,M_{\star}}$ \& $R_{1/2,\,\rm{SFR}}<0.5\,\rm{kpc}$), and dust mass ($R_{1/2,\,M_{\rm{dust}}}\sim1\,\rm{kpc}$), compared to other snapshots at similar redshifts.}
\label{fig:fig_a}
\end{subfigure} 
\begin{subfigure}[b]{1.0\textwidth}
\includegraphics[scale=0.67]{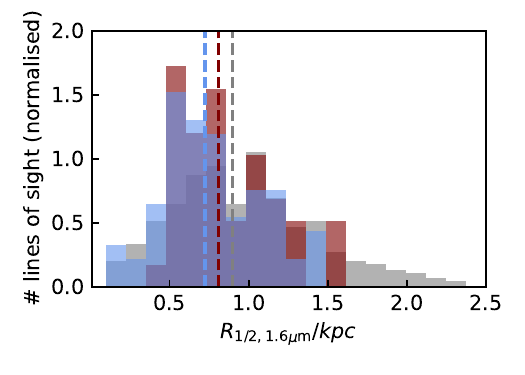}
\includegraphics[scale=0.67]{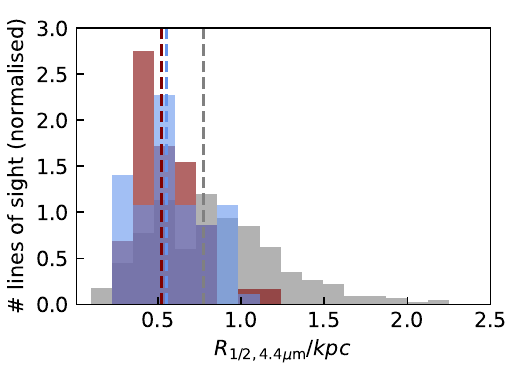}
\includegraphics[scale=0.67]{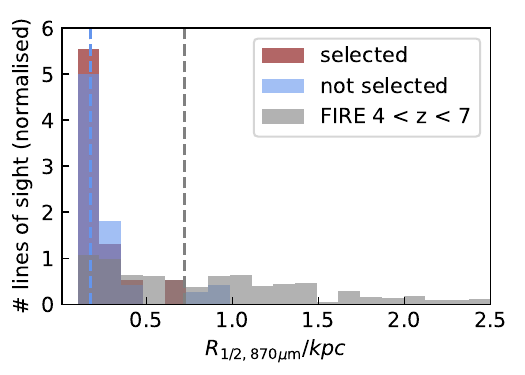}    \caption{Distributions of half-light sizes at observed-frame $1.6\,\mu\rm{m}$ (left), $4.4\,\mu\rm{m}$ (center), and $870\,\mu\rm{m}$ (right) for subsamples of selected and not-selected sightlines (red and blue), compared to those of all FIRE-2 snapshots at $4<z<7$ (grey). Differences in the distributions of half-light radii of selected and not-selected sightlines of snapshots with {\it{HST-}}dark sightlines are minor. {\it{HST-}}dark snapshots typically display more compact half-light radii than others at the same redshift at all three wavelengths, with the largest difference at $870\,\mu\rm{m}$. The half-light sizes (at $1.6\,\mu\rm{m}$ and $4.4\,\mu\rm{m}$) of {\it{HST-}}dark snapshots and the general population are more similar than the half-mass and half-SFR radii. This is because, for the more highly attenuated sources, more of the $1.6\,\mu\rm{m}$ and $4.4\,\mu\rm{m}$ emission is biased to larger radii due to central dust obscuration.}\label{fig:fig_b}
\end{subfigure}
\vspace{-0.5cm}
\caption{Physical (top panel) and observable (bottom panel) radii of {\it{HST-}}dark galaxies compared to all FIRE snapshots within a similar redshift range. }
\label{fig:fig_sizes}
\end{figure*}

\subsection{Physical and light-based sizes of HST-dark sources}\label{sec:HSTfaint_sizes}
\subsubsection{Physical sizes of HST-dark sources}\label{sec:physical_size}
In Figure \ref{fig:fig_a}, we present distributions of the half-mass, half-SFR and half-dust mass sizes of galaxies with and without {\it{HST}}-dark sightlines. These sizes are calculated from particle data, and are hence mass-weighted rather than light-weighted. The stellar mass (both of all stars and recently-formed stars) is extremely compact for the {\it{HST}}-dark sources, with $R_{1/2,\,M_{\star}}$ and $R_{1/2,\,\rm{SFR_{10\,Myr}}}<0.5\,\rm{kpc}$. Although more extended than the stellar mass distribution, the dust mass is also compact ($R_{1/2,\,M_{\rm{dust}}}\sim1\,\rm{kpc}$). For the whole sample at $z=4-7$, the median $R_{1/2,\,\rm{SFR_{10\,Myr}}}/R_{1/2,\,M_{\rm{dust}}}=0.49$. For the subsample of sources with one or more lines of sight meeting the {\it{HST}}-dark criteria, the median $R_{1/2,\,\rm{SFR_{10\,Myr}}}/R_{1/2,\,M_{\rm{dust}}}=0.14$, with median $R_{1/2,\,\rm{SFR_{10\,Myr}}}=0.12\,\rm{kpc}$. Thus, our simulated {\it{HST}}-dark sources feature particularly compact star formation embedded within a more extended dust mass distribution. This drives large optical depths.

\subsubsection{Multi-wavelength sizes of HST-dark sources}\label{sec:half_light_size}
In Figure \ref{fig:fig_b}, we show histograms of half-light radii at $1.6\,\mu\rm{m}$, $4.4\,\mu\rm{m}$ and $870\,\mu\rm{m}$, for selected and not-selected subsamples, as well as for all snapshots in the redshift range $4<z<7$. Note that these half-light sizes are calculated using a curve-of-growth technique, rather than from S\'{e}rsic profile fits to PSF-convolved images (as was performed in \citealt{Parsotan2021,Cochrane2023b}). It is immediately clear that both selected and not-selected lines of sight of {\it HST-}dark snapshots are more compact, on average, than other simulated snapshots within the same redshift range. At $1.6\,\mu\rm{m}$, median half-light radii are $0.81\,\rm{kpc}$ and $0.72\,\rm{kpc}$ for selected and not-selected lines of sight, respectively, compared to $0.90\,\rm{kpc}$ for the whole $4<z<7$ sample. At $4.4\,\mu\rm{m}$, median half-light radii are $0.52\,\rm{kpc}$ and $0.55\,\rm{kpc}$ for selected and not-selected lines of sight, respectively, compared to $0.77\,\rm{kpc}$ for the whole $4<z<7$ sample. The compact sizes of our simulated sources are in line with recent observations: \cite{Gomez-Guijarro2023} found that the F444W-derived effective radii of optically-dark/faint galaxies are 30\% smaller than the average star-forming galaxy, at fixed stellar mass and redshift, with most $\lesssim 1.2\,\rm{kpc}$. The difference between the spatial extent of the $4.4\,\mu\rm{m}$ and $1.6\,\mu\rm{m}$ emission is greatest for the {\it HST-}dark snapshots; this is driven by preferential dust attenuation in the inner regions of the galaxy affecting shorter wavelength light most strongly \citep[see][]{Cochrane2023b} and is in qualitative agreement with the recent observational study of \cite{Suess2022a}.\\
\indent The most significant differences in size are seen at longer wavelengths. At $870\,\mu\rm{m}$, median half-light radii are $0.17\,\rm{kpc}$ for both selected and not-selected lines of sight, compared to $0.72\,\rm{kpc}$ for the whole $4<z<7$ sample. The particularly compact dust continuum emission of observed $K-$dark SMGs was also noted by \cite{Smail2020}. In our simulations, the compact emission is not only driven by a compact dust mass distribution (see Figure \ref{fig:fig_a}, right-hand panel) but by particularly compact star formation (see Figure \ref{fig:fig_a}, central panel; this effect was also discussed in \citealt{Cochrane2019}, who showed that compact dust emission is driven by steep dust temperature gradients associated with compact star formation).

\subsection{Understanding the orientation-dependence of HST-dark source selection}\label{sec:orientation_dependence}
\begin{figure*} 
\includegraphics[width=0.85\columnwidth]{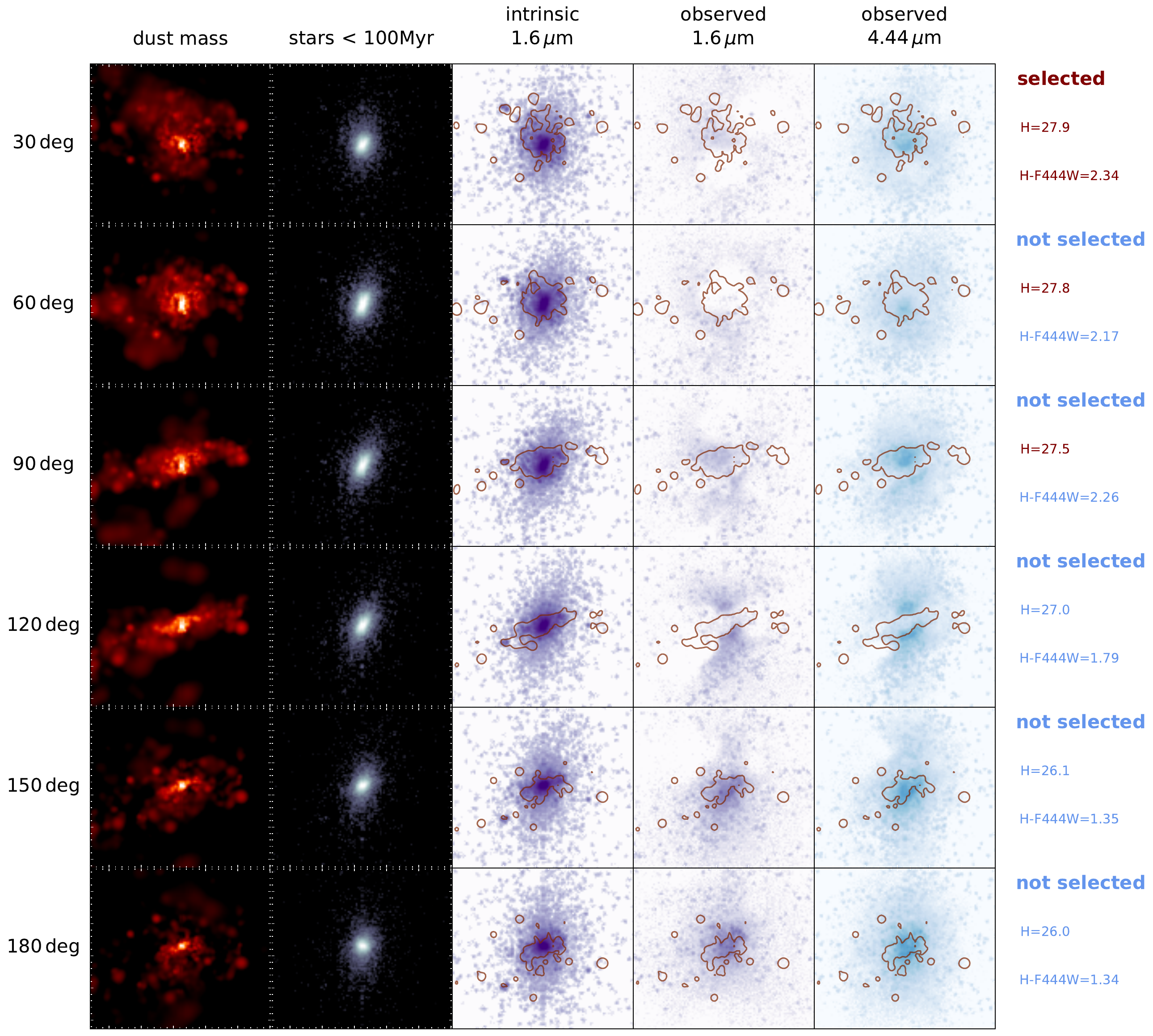}
\includegraphics[width=0.5\columnwidth]{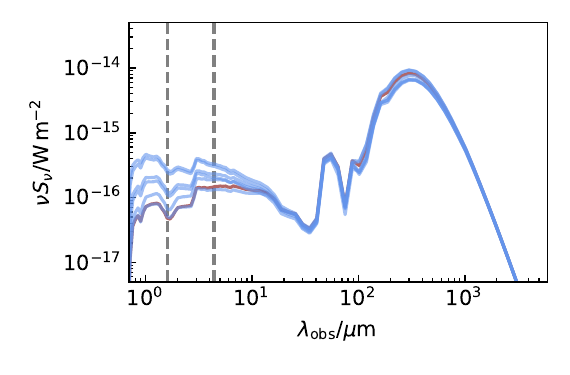}
\includegraphics[width=0.43\columnwidth]{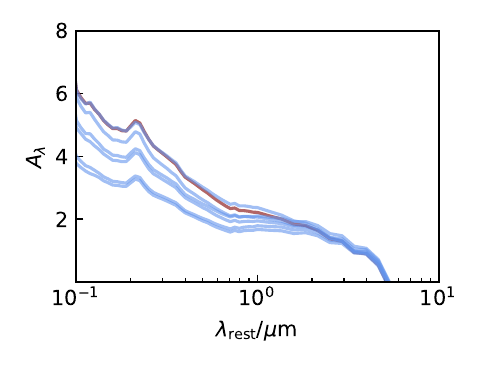}
\caption{Top panels: Projected distributions of dust mass, young stars, and predicted emission (intrinsic and observed) along different lines of sight for galaxy C1 at $z=6.5$. Bottom panels: SED and attenuation curve for the different lines of sight. Vertical dashed lines indicate $H-$band and $4.4\,\mu\rm{m}$. Only one of the simulated lines of sight meets both $H-$band and color selection criteria of {\it HST}-dark galaxies. Two others meet the $H-$band criterion but not the color criterion. Along other lines of sight, the dust does not cover the young stellar emission as completely; hence, they appear brighter in the $H$-band and are also too blue.}
\label{fig:C1_6pt5}
\end{figure*}

\begin{figure*} 
\includegraphics[width=0.85\columnwidth]{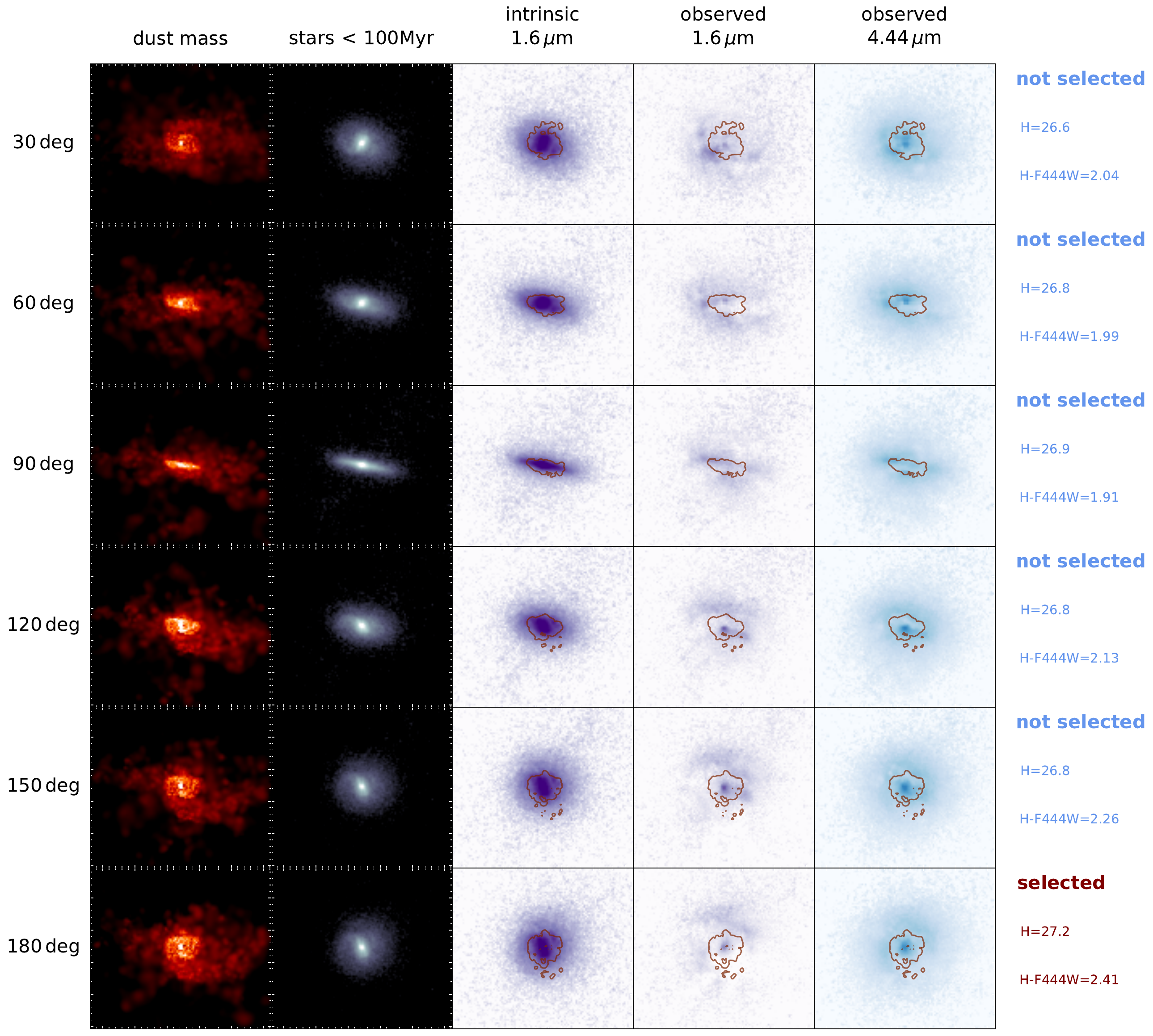}
\includegraphics[width=0.5\columnwidth]{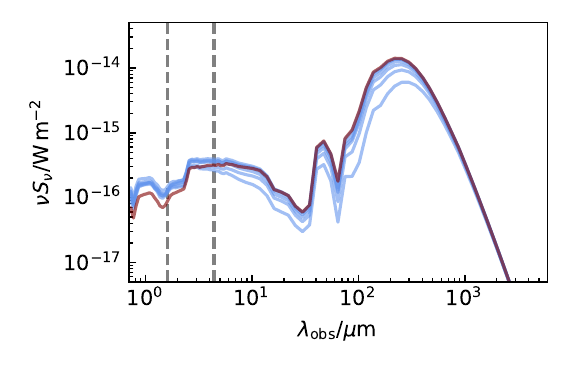}
\includegraphics[width=0.43\columnwidth]{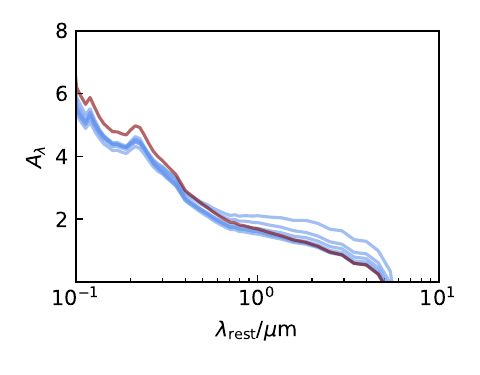}
\caption{The same as Fig \ref{fig:C1_6pt5}, for galaxy B2 at $z=5.4$. Only one line of sight meets the {\it HST}-dark criteria; the others are brighter in $H$ and hence too blue.}
\label{fig:B2_5pt4}
\end{figure*}

In Section \ref{sec:HST-dark_selection}, we noted that for a given simulated galaxy snapshot, there exist lines of sight that meet the $H-$band magnitude and color criteria and others that do not. To illustrate this, in Figures \ref{fig:C1_6pt5} and \ref{fig:B2_5pt4} we show two examples of snapshots selected as {\it HST-}dark along just one of the modelled lines of sight. We plot projected dust mass, recently-formed stars (with age $<100\,\rm{Myr}$), and predicted emission maps at observed-frame $1.6\,\mu\rm{m}$ and $4.4\,\mu\rm{m}$. Contours of the projected dust mass are shown on both emission maps. We also show predicted SEDs and dust attenuation curves for all lines of sight. \\
\indent Figure \ref{fig:C1_6pt5} shows galaxy C1 at $z=6.5$. This galaxy is faintest in the observed-frame $H-$band in the approximately face-on orientation (top panel). This is the only orientation for which both criteria are met. Viewing the galaxy from other angles (subsequent panels), the dust is less well-orientated to cover the emission from the young stars; hence, the galaxy becomes brighter in $H$, which also leads to a bluer $H-\rm{F444W}$ color. Figure \ref{fig:B2_5pt4} shows galaxy B2 at $z=5.4$. This galaxy meets the criteria along just one of the modelled lines of sight; from all others, the galaxy appears too bright in $H$ and also too blue. In both examples, there are lines of sight from which the $H-$band emission appears to escape from the edges of the dust distribution. This is in line with observations of dusty high-redshift galaxies \citep[e.g.][]{Hodge2016,Cochrane2021}.

\subsection{Implications of line-of-sight variations for comparison with models}\label{sec:model_comparison}
\begin{figure} 
\includegraphics[width=\columnwidth]{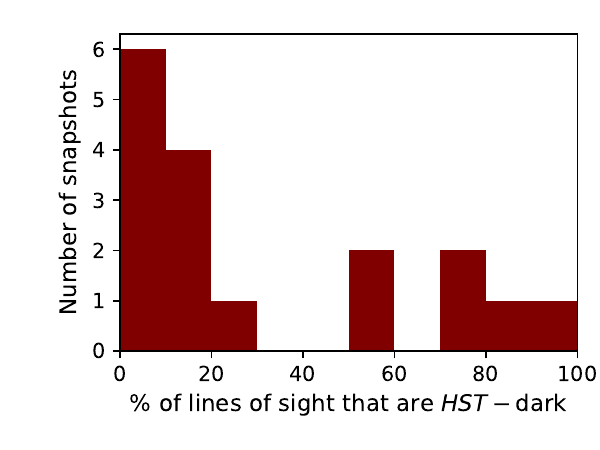}
\caption{We re-ran the radiative transfer procedure for each snapshot with one or more previously-identified {\it{HST}}-dark lines of sight. We used 50 lines of sight, evenly sampling solid angle. Here, we show the distribution of the percentage of snapshots that meet the {\it{HST}}-dark selection criteria. This ranges from $2$ to $90$ per cent.}
\label{fig:number_hist}
\end{figure}

One key result of this paper is the clear dependence of {\it HST-}dark source selection on line of sight. Only one of the 17 selected {\it HST-}dark galaxies is selected from all seven modelled orientations; the rest are selected along some lines of sight but not along others. To investigate this further, we repeat the radiative transfer calculation for each of the $17$ selected snapshots, this time modelling $50$ lines of sight, with solid angle evenly sampled. The number of lines of sight meeting the {\it HST-}dark selection criteria ranges from $1$ to $45$ (see Figure \ref{fig:number_hist}). In total, $268$ of $850$ modelled sightlines meet the criteria ($32\%$). This has implications for the measurement of number densities of {\it{HST}}-dark sources from observations: for each galaxy selected as {\it{HST}}-dark, there are, on average, two others with the same physical properties that would not meet the selection criteria, simply because of sky orientation. More massive, sub-millimeter-bright sources, not simulated here, may be obscured along more lines of sight. \\
\indent The high resolution of the FIRE simulations is key in enabling this result. In coarser resolution simulations, resolving the detailed geometry of gas and stars might not be possible. For example, the baryonic particle resolution of IllustrisTNG-100 is $\sim40$ times lower than our highest resolution simulations \citep{Pillepich2018}, and that of SIMBA's largest box is $\sim14$ times lower \citep{Dave2019}. Furthermore, simulations that do not explicitly resolve stellar feedback (e.g. using effective equation of state models) like IllustrisTNG will artificially suppress small-scale ISM clumping. \\
\indent Semi-analytic models tend to model galaxy geometry in a simplified manner. For example, \cite{Wang2019a} compared the relative contributions of observed $H-$dropouts and modelled $H-$dropouts in the {\sc l-galaxies} SAM (2015 version; \citealt{Henriques2015}) to the cosmic SFRD. However, the SAM models stars only as bulge, disc and intracluster light components. Optical depths are calculated separately for the diffuse ISM and molecular birth clouds, following \cite{DeLucia2007}. The overall extinction curve is derived by assuming an `inclined slab' geometry implementation for the diffuse ISM. Although some geometric effects will be included as a result of the inclination angle, this is a simplistic implementation that does not allow for a clumpy ISM (or scattering into the line-of-sight). We caution that coarse-resolution hydrodynamical simulations and semi-analytic models may not resolve the crucial dust-star geometry sufficiently for robust comparisons with observed {\it HST-}dark populations.\\

\section{Conclusions}\label{sec:conclusions}
Using highly-resolved zoom-in simulations from the FIRE suite, we perform the first detailed modelling of {\it{HST}}-dark galaxies, a population of galaxies identified observationally by their faint $H-$band magnitudes and red $H-\rm{F444W}$ colors. The high resolution of these simulations enables us to infer realistic distributions of dust and stars for galaxies spanning a range of stellar masses and redshifts. We run radiative transfer calculations on eight massive galaxies across $1\lesssim z\lesssim 8$, and make predictions for the observed-frame UV-FIR emission. Applying selection criteria identical to those used in observational searches for IRAC/JWST-bright {\it HST}-dark galaxies, we naturally recover $17$ galaxy snapshots that meet {\it{HST}}-dark selection criteria at one or more viewing angles, all in the redshift range  $z=4-7$. The four most massive halos all pass through a {\it{HST}}-dark stage at some point in their evolution. This is because those halos build enough stellar and dust mass early enough to meet the colour selection.
We then study the physical properties of these galaxies in light of the selections applied. Galaxy snapshots selected as {\it{HST}}-dark show high levels of dust attenuation ($2<A_{V}<4$) compared to others in the same redshift range. Physically, these high $A_{V}$ values are associated with substantial dust masses ($M_{\rm{dust}}>5\times10^{6}\,\rm{M_{\odot}}$) and compact star formation ($R_{\rm{SFR\,10\,Myr}}\sim0.1\,\rm{kpc}$). Modelling the observable emission along additional lines of sight for each snapshot enables us to test the role of geometry in {\it{HST}}-dark galaxy selection. Importantly, galaxies that are {\it{HST}}-dark along some sightlines do not meet the criteria along others. We infer that the observational selection of {\it{HST}}-dark galaxies is subject to a strong viewing angle dependence: rather than a particularly special population, {\it{HST}}-dark galaxies are a subset of high-redshift galaxies viewed along lines of sight with particularly high dust attenuation. This result has implications for comparisons of observations with semi-analytic models and coarse-resolution simulations that do not resolve the detailed geometry of stars and dust. 

\section*{Acknowledgements}
We thank the anonymous reviewer for helpful suggestions on an early version of this paper. The Flatiron Institute is supported by the Simons Foundation. The simulations presented in this work were run on the Flatiron Institute’s research computing facilities (Gordon-Simons, Popeye, and Iron compute clusters), supported by the Simons Foundation. DAA acknowledges support by NSF grants AST-2009687 and AST-2108944, CXO grant TM2-23006X, Simons Foundation Award CCA-1018464, and Cottrell Scholar Award CS-CSA-2023-028 by the Research Corporation for Science Advancement. FC acknowledges support from a UKRI Frontier Research Guarantee Grant (PI Cullen; grant reference EP/X021025/1).
\section*{Data availability}
The FIRE-2 simulations are publicly available \citep{Wetzel2022} at \url{http://flathub.flatironinstitute.org/fire}. Additional FIRE simulation data is available at \url{https://fire.northwestern.edu/data}.
A public version of the \textsc{Gizmo} code is available at \url{http://www.tapir.caltech.edu/~phopkins/Site/GIZMO.html}.

\bibliographystyle{aasjournal}
\bibliography{Edinburgh}

\end{document}